\documentclass[proof]{WileyASNA-v1}

\usepackage{xcolor}

\articletype{Article Type}%

\received{26 April 2016}
\revised{6 June 2016}
\accepted{6 June 2016}

\begin{document} 

\title{Cold Magnetized Quark Matter under Compact-Star Conditions}

\author[1,2]{S.A. Ferraris*}

\author[3,4]{J.P. Carlomagno}

\author[3,4]{G.A. Contrera}

\author[4,5]{A.G. Grunfeld}    
\authormark{FERRARIS \textsc{et al}}

\address[1]{  
    {Facultad de Ingeniería y Ciencias Agrarias, Pontificia Universidad Católica Argentina,}\\
    { Av. Alicia Moreau de Justo 1500, (C1107AAZ) Buenos Aires, Argentina.}  }
\address[2]{ {{ Departamento de Física Teórica de Interacciones Fundamentales y Sistemas Complejos, Laboratorio Tandar,}\\
    { Comisión Nacional de Energía Atómica,} \\
    { Av. del Libertador 8250, (C1429BNP) Buenos Aires, Argentina.} } }
\address[3]{{IFLP, CONICET departamento de Física, Facultad de Ciencias Exactas, Universidad Nacional de La Plata, C.C. 67, (1900) La Plata, Argentina.}}    

\address[4]{{CONICET, Godoy Cruz 2290, Buenos Aires, Argentina.}}

\address[5]{{Instituto de Astronomía y Física del Espacio (IAFE, CONICET-UBA), Intendente Güiraldes 2160 - Ciudad Universitaria, Buenos Aires, Argentina.} }

\corres{*Av. General Paz, 1499 - San Mart\'in, Buenos Aires, B1650KNA.  \email{ferrarissebastian@uca.edu.ar}}


\abstract{We investigate the phase structure of cold magnetized two-flavor quark matter under compact-star conditions within a nonlocal chiral quark model with separable interactions. Electric charge neutrality and beta equilibrium are imposed in order to describe matter under realistic neutron-star conditions, while nonlocality is implemented through a Gaussian form factor. We find that low and moderate magnetic fields reduce the critical chemical potential for chiral symmetry restoration, signaling inverse magnetic catalysis. For stronger magnetic fields, the critical chemical potential becomes nearly independent of the magnetic field strength,  Compared with the isospin-symmetric case, the combined effects of charge neutrality and beta equilibrium shift the first-order chiral transition to slightly higher chemical potentials, indicating that compact-star constraints delay the onset of chiral symmetry restoration. We further show that charge neutrality and beta equilibrium modify the de Haas--van Alphen oscillations in the quark density, shifting the Landau-level thresholds for the $u$ and $d$ quarks in opposite directions relative to the isospin-symmetric case. The effect on the chiral transition is small ($\sim 5$~MeV) for fields $B\lesssim 10^{19}$~G, relevant for magnetar interiors, and becomes sizable only for $B\gtrsim 3\times10^{19}$~G. These results highlight the importance of incorporating realistic stellar conditions in studies of the QCD phase diagram at high baryon density.
}




\maketitle


\section{INTRODUCTION}\label{sec1}

The properties of strongly interacting matter in the presence of intense magnetic fields have attracted considerable attention in recent years, particularly in connection with the physics of compact stars. While surface magnetic fields in magnetars are of the order of $10^{15}\,\mathrm{G}$, fields as large as $10^{18}$--$10^{19}\,\mathrm{G}$ may be reached in their interiors, the upper bound being set by virial-theorem arguments~\cite{LaiShapiro:1991,Chakrabarty:1997}. In natural units, $eB = 1$~GeV$^2$ corresponds to $B\simeq 1.69\times10^{20}$~G, so that this astrophysically relevant range translates into $eB\lesssim 0.06$~GeV$^2$.  Under these extreme conditions, matter is expected to satisfy electric charge neutrality and beta equilibrium, making compact stars a unique laboratory for exploring the QCD phase diagram at high baryon densities.

Since first-principles QCD calculations at finite baryon chemical potential are still hindered by the sign problem, effective models remain indispensable for investigating strongly interacting matter in this regime. Among them, nonlocal chiral quark models provide a realistic description of low-energy QCD by incorporating momentum-dependent interactions and reproducing several features observed in lattice QCD calculations. In particular, previous studies have shown that within this framework inverse magnetic catalysis arises without introducing an explicit magnetic-field dependence in the model parameters~\cite{Pagura:2017}.
Beyond their role in the chiral transition, de Haas--van Alphen oscillations are also of phenomenological interest, as they may influence the magnetization and stability of
magnetized quark matter in compact-star interiors \cite{Noronha:2007,Huang:2010}.

The phase structure of magnetized quark matter under isospin-symmetric conditions has been previously investigated within the same framework \cite{Pagura:2017,Ferraris:2021vun}. In the present work, we extend this analysis by imposing the conditions relevant for matter inside compact stars, namely electric charge neutrality and beta equilibrium. This enables us to isolate and quantify the impact of these constraints on the chiral phase transition and the phase structure of dense magnetized quark matter. We show that, compared with the isospin-symmetric case, compact-star conditions shift the first-order chiral transition toward slightly higher chemical potentials, indicating a delayed onset of chiral symmetry restoration.

\section{MODEL}\label{sec2}

We consider cold two-flavor quark matter described by a covariant nonlocal chiral quark model in the presence of a uniform external magnetic field,
$\mathbf{B}=B\,\hat{z}$. The formalism has been presented in detail in
Refs.~\cite{Pagura:2017,GomezDumm:2017,GomezDumm:2021}; here we briefly summarize the ingredients relevant for the present work.

We work in the chiral limit, $m_u=m_d=0$, and employ a Gaussian nonlocal form factor
\begin{equation}
g(p^2)=\exp\left(-p^2/\Lambda^2\right),
\end{equation}
where $\Lambda$ denotes the nonlocality scale. 
Regarding the nonlocal interaction, the Gaussian form factor $g(p^2) = \exp(-p^2/\Lambda^2)$ in Euclidean momentum space corresponds in coordinate space to a separable interaction kernel. This regularizes the ultraviolet behavior of the theory in a soft, non-abrupt manner and endows the quark self-energy with a dynamically generated momentum-dependent constituent mass $M(p)$. The local limit is recovered only when $\Lambda \to \infty$, where the interaction kernel collapses into a Dirac delta function. While other non-Gaussian functional forms (such as Lorentzian or exponential profiles) can be considered, the qualitative features of the chiral transition remain largely form-factor independent \cite{GomezDumm:2006,Pagura:2017}, and the Gaussian choice enables the transverse momentum integrations in the presence of magnetic fields to be carried out analytically.

Explicitly, the nonlocal interaction is built from the current
$j(x)=\int d^4z\,\mathcal{G}(z)\,\bar\psi(x+z/2)\,\psi(x-z/2)$, where
$\mathcal{G}(z)$ is the Fourier transform of $g(p^2)$. Hence, although the
Gaussian form factor provides a fast UV convergence, the quark fields
interact over a finite space--time region rather than at a single point,
i.e., the interaction does not reduce to a contact one but has a well-defined finite range.
In the presence of the external field, gauge invariance requires the
fields at different points to be connected by a Wilson line,
$\psi(x)\to W(x,s)\,\psi(s)$, which gives rise to Schwinger-type phases.
Using the Ritus eigenfunction method, the mean-field quark propagator becomes
diagonal in the Landau-level basis, and the dynamical mass acquires a
dependence on the Landau level $k$, the flavor charge, and the parallel
momentum, $M^{\lambda,f}_{k}(p_\parallel)=\bar\sigma\,g^{\lambda,f}_{k}(p_\parallel)$
\cite{Pagura:2017,GomezDumm:2017}. For the Gaussian form factor one obtains

\begin{equation}
g^{\lambda,f}_{k}(p_\parallel)=
\frac{\left(1-|q_fB|/\Lambda^2\right)^{k_\lambda}}
{\left(1+|q_fB|/\Lambda^2\right)^{k_\lambda+1}}\,
e^{-p_\parallel^2/\Lambda^2}.
\label{gk}
\end{equation}
Therefore, the strength of the nonlocal effects in the magnetized medium is
governed by the ratio $|q_fB|/\Lambda^2$, i.e., by the magnetic length
$\ell_f=|q_fB|^{-1/2}$ compared with the nonlocality range $1/\Lambda$.
This Landau-level dependence of the effective mass, absent in local
NJL-type models, is the mechanism responsible for inverse magnetic
catalysis within this framework, without any $B$-dependent coupling.

The model parameters are taken from Ref.~\cite{GomezDumm:2006}. Within the mean-field approximation, translationally invariant scalar mean fields and vanishing pseudoscalar mean fields are assumed. Since the scalar interaction acts only in the isoscalar channel, a single flavor-independent mean field can be taken also under compact-star conditions, where $\mu_u\neq\mu_d$ and $q_u\neq q_d$:
\begin{equation}
\bar{\sigma}_u=\bar{\sigma}_d\equiv\bar{\sigma},
\end{equation}
where $\bar{\sigma}$ acts as the order parameter for chiral symmetry breaking.

The regularized thermodynamic potential is given by
\begin{equation}
\Omega^{\rm reg}
=
\frac{\bar{\sigma}^{\,2}}{2G}
+\sum_{f=u,d}\Omega_f^{\rm reg}
+\Omega_e,
\end{equation}
where $\Omega_f^{\rm reg}$ denotes the regularized quark contribution for flavor $f=u,d$, while $\Omega_e$ is the electron contribution. The explicit expressions for $\Omega_f^{\rm reg}$ are given in Refs.~\cite{Pagura:2017,GomezDumm:2017,GomezDumm:2021} and are not repeated here.

To describe matter under compact-star conditions, beta equilibrium and electric charge neutrality are imposed self-consistently. The flavor chemical potentials are related to the quark and electron chemical potentials according to
\begin{equation}
\mu_u=\mu-\frac{2}{3}\mu_e,
\qquad
\mu_d=\mu+\frac{1}{3}\mu_e,
\end{equation}
which satisfy the beta-equilibrium condition
\begin{equation}
\mu_d=\mu_u+\mu_e.
\end{equation}

The electric charge neutrality condition reads
\begin{equation}
\frac{2}{3}n_u-\frac{1}{3}n_d-n_e=0,
\label{charge_neut}
\end{equation}
where the quark number densities are obtained from
\begin{equation}
n_f=
-\frac{\partial\Omega^{\rm reg}}{\partial\mu_f},
\qquad
f=u,d.
\end{equation}

The occupied Landau levels are determined by the condition
\begin{equation}
\mu_f \geq S_{kf},
\qquad
S_{kf}=\sqrt{2k|q_f|B},
\end{equation}
valid for massless quarks in the chirally restored phase,
which implies that the highest populated Landau level is is given by the integer part

\begin{equation}
k_{f,\mathrm{max}}
=
\left\lfloor \frac{\mu_f^2}{2\lvert q_f\rvert B} \right\rfloor ,
\end{equation}

where $\lfloor x \rfloor$ denotes the largest integer smaller than or equal to $x$.
Consequently, successive Landau levels become populated whenever
\begin{equation}
\mu_f=\mu_{c,k}^{(f)}
=\sqrt{2k|q_f|B},
\end{equation}
giving rise to the de Haas--van Alphen oscillations observed in several thermodynamic quantities.

For electrons, the density is given by
\begin{equation}
n_e=
\frac{eB}{2\pi^2}
\sum_{k=0}^{k_{e,\rm max}}
\alpha_k
\sqrt{\mu_e^2-2keB},
\end{equation}
where
\begin{equation}
k_{e,\rm max} = \left\lfloor\frac{\mu_e^2}{2eB} \right\rfloor  ,
\end{equation}
where $\alpha_0=1$ and $\alpha_{k>0}=2$ account for the spin degeneracy of the Landau levels, and the electron mass has been neglected.

The equilibrium state is determined by solving self-consistently the gap equation
\begin{equation}
\frac{\partial\Omega^{\rm reg}}
{\partial\bar{\sigma}}=0,
\end{equation}
together with the charge-neutrality condition, Eq.~(\ref{charge_neut}), for fixed values of the quark chemical potential $\mu$ ($\equiv \mu_B/3$) and the magnetic field strength. Throughout this work, all results are presented as functions of $\mu$, while the flavor and electron chemical potentials are obtained self-consistently from the beta-equilibrium and charge-neutrality conditions.

\section{RESULTS}\label{sec3}

The model is characterized by two parameters, namely the nonlocality
scale $\Lambda$ and the coupling constant $G$. Following
Ref.~\cite{GomezDumm:2006}, we adopt the parameter set that
reproduces $f_{\pi,\rm ch}=90$ MeV and
$-\langle\bar q q\rangle_{\rm ch}^{1/3}=230$ MeV, leading to
$\Lambda=684.054$ MeV and $G\Lambda^2=23.42$.
In order to quantify the impact of compact-star constraints on
the phase structure of magnetized quark matter, we compare our
new chiral-limit results for quark matter under compact-star
conditions (CSC) with those previously obtained for isospin-symmetric
matter (ISM) within the same framework, published in
\citet{Ferraris:2021vun}.
Unlike Ref.~\cite{Pagura:2017}, the present analysis incorporates
beta equilibrium and electric charge neutrality, as required for
matter in compact-star interiors. For each value of the magnetic field
and quark chemical potential, the equilibrium state is obtained by
solving self-consistently the gap equation together with the
charge-neutrality condition.
To connect with astrophysical scenarios, we recall that
$eB=1$~GeV$^2$ corresponds to $B\simeq1.69\times10^{20}$~G. 

\begin{figure}[htbp]
    \centering
    \includegraphics[width=0.48\textwidth]{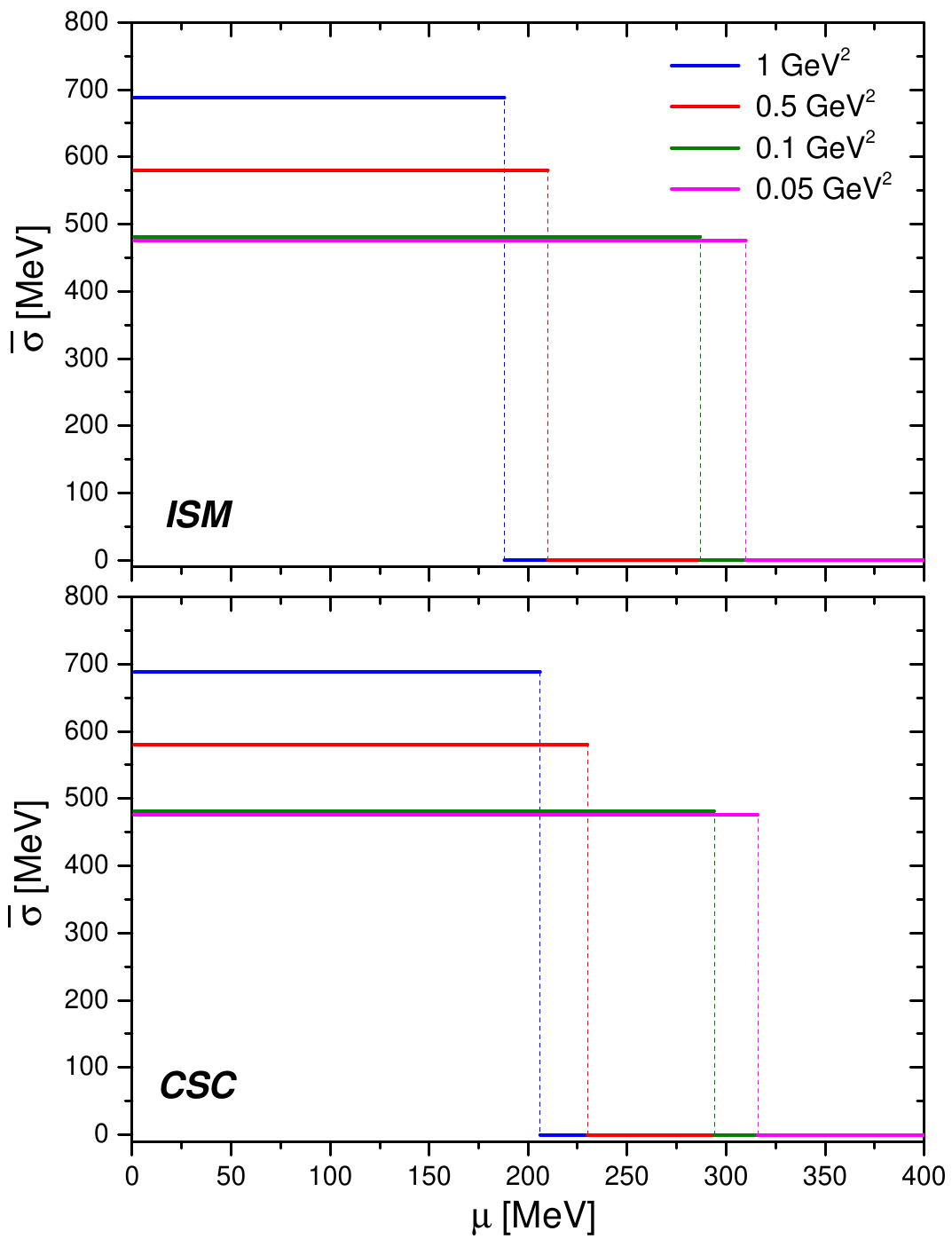}
    \caption{Behavior of the mean field parameter $\bar{\sigma}$ as a function of the quark chemical potential $\mu$ under isospin-symmetric matter (ISM, top panel) and compact-star conditions (CSC, bottom panel). The curves correspond to selected values of the external magnetic field $eB$: $1$, $0.5$, $0.1$, and $0.05 \text{ GeV}^2$ ($B\simeq1.7\times10^{20}$, $8.4\times10^{19}$, $1.7\times10^{19}$ and $8.4\times10^{18}$~G, respectively)}. 
    \label{figura1}
\end{figure}

\begin{figure}[htbp]
    \centering
    \includegraphics[width=0.48\textwidth]{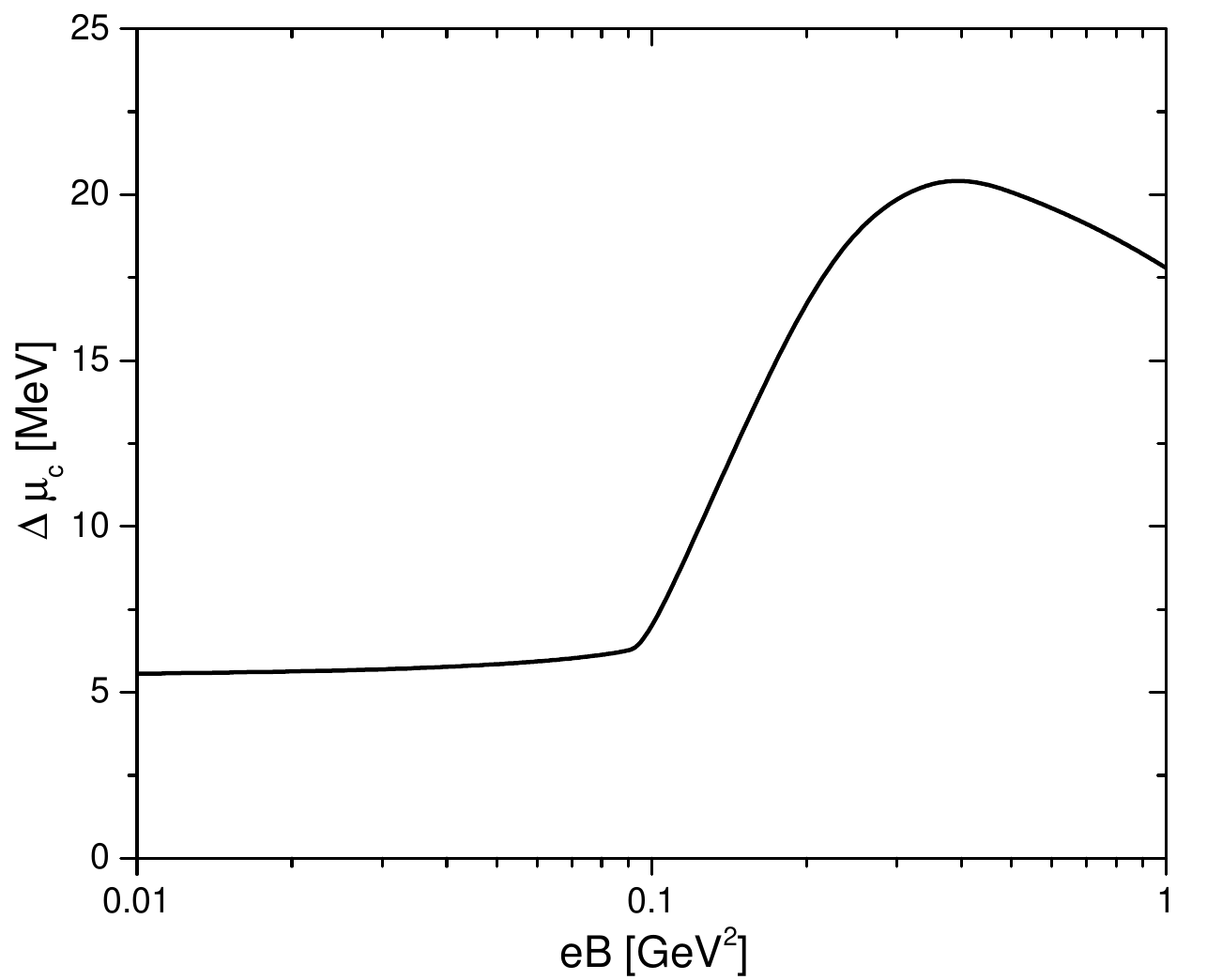}
    \caption{Difference in the critical chemical potential, $\Delta\mu_c=\mu_c^{\text{CSC}} - \mu_c^{\text{ISM}}$, between compact-star conditions (CSC) and isospin-symmetric matter (ISM) as a function of the magnetic field $eB$. }
    \label{figura2}
\end{figure}

\begin{figure}[htbp]
    \centering
    \includegraphics[width=0.48\textwidth]{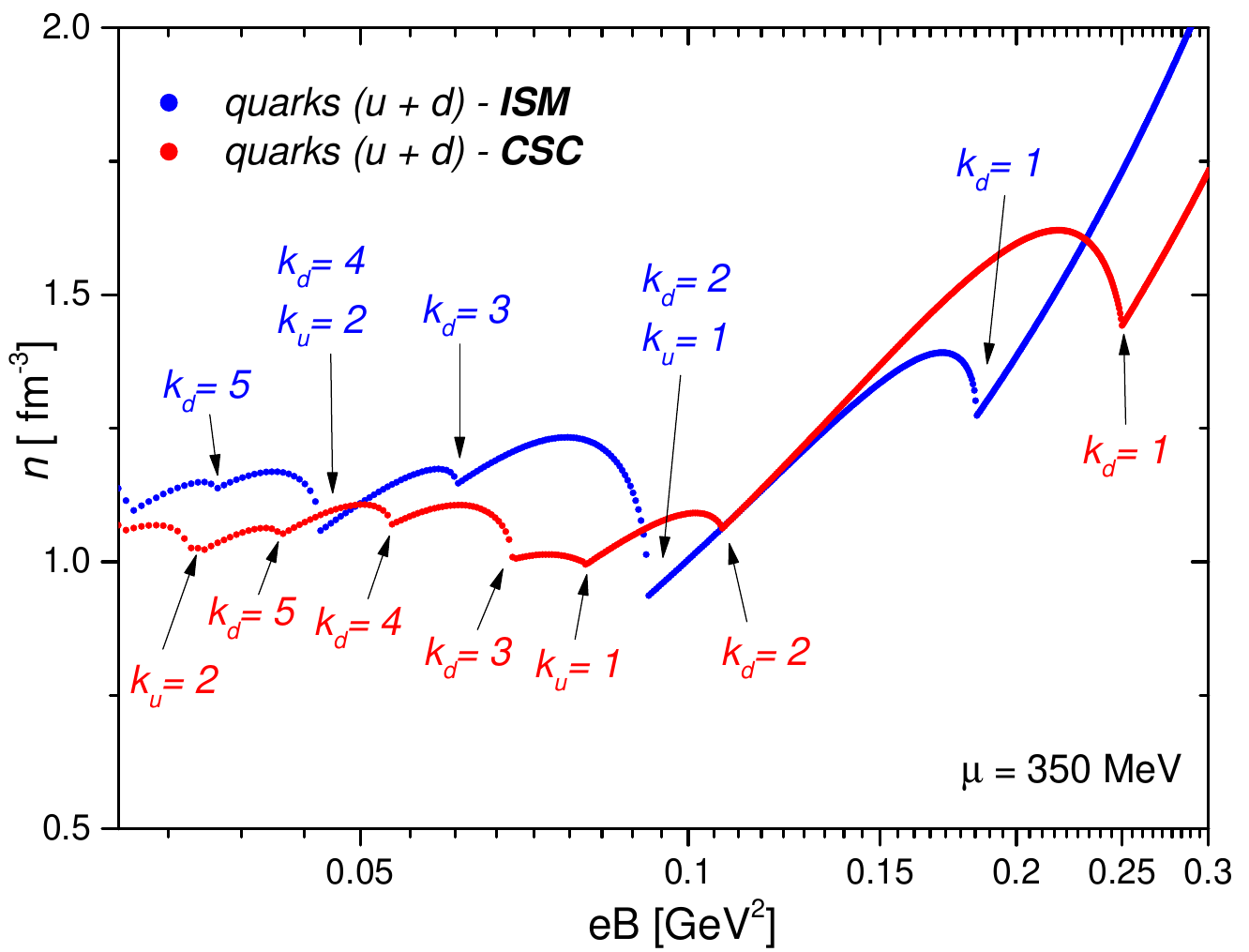}
    \caption{{Behavior of the total quark density $n$ vs. the magnetic field $eB$ at $\mu = 350~\text{MeV}$ for ISM (blue) and CSC (red) configurations. The arrows indicate the sequential (de)population of $u$ and $d$ quark Landau levels (labeled as $k_u$ and $k_d$), which give rise to the characteristic de Haas--van Alphen} oscillations.} 
    \label{figura3}
\end{figure}

Figure~\ref{figura1} shows the behavior of the chiral order parameter
$\bar{\sigma}$ as a function of the quark chemical potential for
selected values of the magnetic field. In all cases, the order
parameter exhibits a discontinuous jump, indicating that the chiral
phase transition remains of first order. Comparing the upper and lower
panels, one observes that the combined effect of beta equilibrium and
electric charge neutrality systematically shifts the transition toward
slightly larger values of the quark chemical potential. Therefore,
compact-star conditions delay the onset of chiral symmetry restoration
with respect to the isospin-symmetric case, while preserving the
qualitative features of the transition.
Moreover, in both cases $\mu_c$ decreases with increasing magnetic field: under CSC it goes from $\mu_c\simeq317$~MeV at $eB=0.05$~GeV$^2$ to $\simeq295$, $231$ and $207$~MeV at $eB=0.1$, $0.5$ and $1$~GeV$^2$, respectively, which signals inverse magnetic catalysis.

To quantify the effect of the compact-star constraints on the chiral
transition, we define
\begin{equation}
\Delta\mu_c(eB)
=
\mu_c^{CSC}(eB)
-
\mu_c^{{ISM}}(eB),
\label{delta_muc}
\end{equation}
where $\mu_c^{ISM}$ denotes the critical chemical potential
obtained in isospin-symmetric quark matter. The resulting difference
is shown in Fig.~\ref{figura2}.

Interestingly, $\Delta\mu_c$ remains positive over the whole range
of magnetic fields considered, showing that the simultaneous
imposition of beta equilibrium and electric charge neutrality
systematically delays chiral symmetry restoration. The magnitude of this effect, however, exhibits a strongly nontrivial dependence on the magnetic field. For $eB\lesssim0.1$~GeV$^2$ ($B\lesssim1.7\times10^{19}$~G),
$\Delta\mu_c$ remains small and nearly constant ($\sim5$--$6$~MeV),
indicating that in this regime the chiral transition is only weakly affected by charge neutrality and beta equilibrium. Notably, this is the regime of direct relevance for magnetar interiors. Beyond $eB\sim0.1$~GeV$^2$, $\Delta\mu_c$ rises sharply, reaching
a maximum of approximately 20~MeV around $eB\sim0.3$--$0.4$~GeV$^2$($B\simeq5$--$7\times10^{19}$~G),
before decreasing gradually to about 18~MeV at $eB=1$~GeV$^2$ ($B\simeq1.7\times10^{20}$~G).

This behavior indicates that the impact of compact-star constraints on the chiral transition is concentrated at intermediate
and strong magnetic fields, where the population of Landau
levels is reduced to a small number of levels approaching the
lowest Landau level (LLL). In this regime, the flavor-dependent
splitting of the chemical potentials induced by charge neutrality
has the largest effect on the location of the critical chemical
potential, while at weaker fields, where many Landau levels
are populated for both quark flavors, this splitting has a comparatively small effect on $\mu_c$.

This behavior shows that the effect of the compact-star constraints
cannot be represented by a simple constant shift of the critical
chemical potential. Instead, the interplay between the magnetic field,
the flavor-dependent chemical potentials, and the charge-neutrality
condition leads to a nontrivial field dependence of the chiral
transition. In particular, the largest modification occurs in the
intermediate-field region, where the magnetic field has a substantial
impact on the population of the different Landau levels.

Figure~\ref{figura3} shows the total quark number density, $n=n_u+n_d$, as a function of the magnetic field strength $eB$ (in GeV$^2$; the range shown corresponds to $B\simeq 0.8$--$5\times10^{19}$~G), for two different conditions on the quark chemical potentials at fixed $\mu=350$~MeV: isospin-symmetric matter (ISM, blue), where $\mu_u=\mu_d\equiv\mu$ with no constraint from charge neutrality, and the compact-star condition (CSC, red), where charge neutrality and beta equilibrium are enforced self-consistently together with the electron contribution.

Both curves display the characteristic sawtooth-like de Haas--van Alphen oscillations arising from the successive (de)population of Landau levels as $eB$ is varied: the density rises smoothly until a given level $k$ empties, producing a sharp downward kink, after which growth resumes with a smaller-amplitude tooth. The individual level closures for the $u$ and $d$ quarks are explicitly labeled ($k_u=1,2$ and $k_d=1,\dots,5$); the $d$ quark exhibits a denser sequence of transitions than the $u$ quark over the same $eB$ range, a direct consequence of its smaller electric charge ($|q_d|<|q_u|$), which yields a finer Landau-level spacing $2|q_f|B$.

Comparing the two curves reveals the effect of charge neutrality and beta equilibrium on the oscillation pattern: while ISM and CSC track each other closely at low $eB$, imposing neutrality shifts $\mu_u$ and $\mu_d$ upward relative to the symmetric value $\mu$ (through the nonzero electron chemical potential required to keep the system neutral). This asymmetric shift displaces the de Haas--van Alphen transitions in opposite directions for each flavor: $u$-quark thresholds move to slightly lower $eB$ in the CSC curve, while $d$-quark thresholds move to higher $eB$ --- most visibly for the last, largest-amplitude transition ($k_d=1$), which occurs at $eB\approx0.19$~GeV$^2$ ($B\simeq3.2\times10^{19}$~G) for ISM but is delayed until $eB\approx0.25$~GeV$^2$ ($B\simeq4.2\times10^{19}$~G) for CSC. The two curves cross near $eB\sim0.1$~GeV$^2$ and again near the upper edge of the plotted range, reflecting the cumulative effect of these opposite flavor-dependent shifts on the total density.
A qualitatively similar decoupling of the flavor-dependent de Haas--van Alphen
thresholds under charge neutrality, including transitions that merge and
subsequently split as the chemical potential increases, has been reported
within the local NJL model~\cite{Allen:2015}, suggesting that this behavior
is a generic consequence of charge neutrality.

\section{CONCLUSIONS}\label{sec4}

We have studied the phase structure of cold, two-flavor magnetized quark matter within a nonlocal chiral quark model, imposing electric charge neutrality and beta equilibrium as required by compact-star conditions. The chiral transition remains first order, with the critical chemical potential $\mu_c$ decreasing with $eB$ at low and moderate fields (inverse magnetic catalysis) and becoming nearly field-independent at stronger fields.

Compared with isospin-symmetric matter, compact-star conditions systematically shift $\mu_c$ to higher values, with $\Delta\mu_c=\mu_c^{CSC}-\mu_c^{ISM}$ remaining positive throughout, peaking at $\sim20$~MeV around $eB\sim0.3$--$0.4$~GeV$^2$ ($B\sim5$--$7\times10^{19}$~G). For fields of direct relevance to magnetar interiors, $B\lesssim10^{19}$~G, the shift is moderate, $\Delta\mu_c\sim5$--$6$~MeV. Charge neutrality and beta equilibrium also displace the de Haas--van Alphen oscillations in the quark density, shifting the $u$- and $d$-quark Landau-level thresholds in opposite directions relative to the symmetric case. These results show that realistic stellar conditions have a nontrivial impact on both the chiral transition and the Landau-level structure of dense magnetized quark matter.

\section{Acknowledgments}
This work has been supported in part by Consejo Nacional de Investigaciones Cientificas y Técnicas under Grants No. PIP2022-GI-11220210100150CO and by the National University of La Plata (Argentina), Project No. X960.

\bibliography{referencias}%

\end{document}